\documentclass[aps,pre,twocolumn,floatfix,showpacs,groupaddress]{revtex4-2}

\usepackage{amssymb}
\usepackage{amsmath}
\usepackage{bm}
\usepackage{graphicx}
\usepackage{hyperref}
\usepackage{mathtools}

\usepackage[utf8]{inputenc}
\newcommand{\Q}{\mathbf{Q}}
\newcommand{\n}{\mathbf{\hat{n}}}
\newcommand{\OMEGA}{\bm{\hat{\Omega}}}
\newcommand{\T}{\mathbf{\hat{T}}}

\newcommand{\XI}{\bm{\hat{\Xi}}}
\newcommand{\DN}{\mathbf{D}^{(\hat{\mathbf{n}})}}

\begin{document}
\title{A tensor density measure of topological charge in three dimensional nematic phases}
\author{Cody D. Schimming}
\email{cschim@lanl.gov}
\affiliation{Theoretical Division and Center for Nonlinear Studies, Los Alamos National Laboratory, Los Alamos, New Mexico 87545, USA}

\author{Jorge Vi\~nals}
\affiliation{School of Physics and Astronomy, University of Minnesota, Minneapolis, Minnesota 55455, USA}

\begin{abstract}
   A path independent measure in order parameter space is introduced such that, when integrated along any closed contour in a three dimensional nematic phase, it yields the topological charge of any line defects encircled by the contour. A related measure, when integrated over either closed or open surfaces, reduces to known results for the charge associated with point defects (hedgehogs) or Skyrmions. We further define a tensor density, the disclination density tensor $\mathbf{D}$, from which the location of a disclination line can be determined. This tensor density has a dyadic decomposition near the line into its tangent and its rotation vector, allowing a convenient determination of both. The tensor $\mathbf{D}$ may be nonzero in special configurations in which there are no defects (double-splay or double-twist configurations), and its behavior there is provided. The special cases of Skyrmions and hedgehog defects are also examined, including the computation of their topological charge from $ \mathbf{D}$. 
\end{abstract}

\maketitle

\section{Introduction}

The study of line defects in three dimensional nematic phases, known as disclinations, has gained considerable interest recently following advances in both experimental diagnostics and computational methods \cite{schimming20b,guo21,schimming21,zushi22,wang23,modin23}. Disclinations and their motion have found applications to microfluidics, colloidal self-assembly, surface actuation, optical control, and active and biological matter \cite{ravnik07,copar11,peng15,re:conklin17,saw17,opathalage19,duclos20,zhang21,zhang21b,meng23}. Recent theoretical research has also advanced our understanding of disclinations, including measures of their topology and geometry, as well as analytical results for disclination kinematics \cite{tang17,angheluta21,long21,schimming22,schimming23}.

The well established definition of the topological charge $m$ of a disclination in two dimensional nematics can be expressed as a function of the director angle $\phi$, the director $\n$, or the tensor order parameter $\Q$ \cite{deGennes75}, as
\begin{align}
    m &= \frac{1}{2\pi}\oint_C \partial_k \phi \, d\ell_k \label{eqn:PhiCharge}  \\
    &= \frac{1}{2\pi}\oint_C \varepsilon_{\mu \nu} \hat{n}_{\mu} \partial_k \hat{n}_{\nu} \, d\ell_k \label{eqn:nCharge} \\
    &= \frac{1}{2\pi S_N^2} \oint_{C^*} \varepsilon_{\mu \nu} Q_{\mu \alpha}\partial_k Q_{\nu \alpha} d\ell_k \label{eqn:Qcharge}
\end{align}
where repeated indices are summed over, and $C$ and $C^{*}$ are closed curves encircling the defect, but the latter is restricted to a path far from the core in which the scalar order parameter $S = S_N$ is constant. $\epsilon_{\mu \nu}$ is the Levi-Civita tensor in two dimensions. Stokes' theorem may be applied and the line integrals transformed to surface integrals over surfaces bounded by curves $C$ or $C^*$. The integrands of these surface integrals may then be regarded as locally defined ``densities'', carrying information about the topological charge. For the case of the singular quantities $\phi$ and $\n$, these densities are Dirac delta functions located at cores of defects. For the case of Eq. \eqref{eqn:Qcharge}, this density is a diffuse scalar field with maxima or minima located at defect cores. 

In this work, an invariant measure is introduced such that, when integrated along any closed contour in a three dimensional nematic configuration, it yields the topological charge of any line defects encircled by the contour. A related measure, when integrated over either closed or open surfaces, reduces to known results for the charge associated with point defects (hedgehogs) or Skyrmions.

Generalizing Eqs. \eqref{eqn:PhiCharge}--\eqref{eqn:Qcharge} to three dimensional disclination lines has not been possible due to several added complications. First, the topology of the ground state manifold is different between two and three dimensional nematics. The ground state manifold in three dimensions is the two dimensional real projective space $\mathbb{RP}^2$ which is not isomorphic to the unit sphere, and is non-orientable. This space is equivalent to a hemisphere in which all points on the equator are identified with their polar opposites \cite{alexander12}. This results in all line disclinations having a topological charge of $+1/2$, and any two disclinations that come into contact will annihilate \cite{alexander12}. This is in contrast with two dimensions. There, the ground state manifold is $\mathbb{RP}^1$ which is isomorphic to the unit circle $S^{1}$. Disclinations in two dimensions can have positive and negative charge, and they combine according to well established rules. An invariant measure of charge can be readily defined in two dimensions from closed paths on the circle, Eqs. \eqref{eqn:PhiCharge}--\eqref{eqn:Qcharge}. In three dimensional space, two angles are needed to describe the director orientation. With this added dimensionality, Eqs. \eqref{eqn:PhiCharge}--\eqref{eqn:Qcharge} do not have a direct generalization as a continuum of circuits can be constructed encircling a disclination that lead to a different length in order parameter space. An extension of these expressions to paths on $\mathbb{RP}^2$ has not yet been given, and it is the subject of our work below. 

Second, the geometric structure of disclination lines in three dimensions is quite complex as they can be of wedge, twist, or mixed type \cite{deGennes75,duclos20}. Disclination interactions and motion are governed not just by their topological charge but by their geometrical structure as well. It is well established that the geometry of the disclination line can be characterized by a single vector, the rotation vector $\OMEGA$: The director and its distortion in the vicinity of the disclination lie on the plane defined by $\OMEGA$. Note that $\OMEGA$ itself may change along the disclination line. It is also shown below that the new measure which we introduce contains the local rotation vector as well. 

Finally, three dimensional systems allow biaxiality, and they are commonly described by the tensor order parameter $\Q$. Although $\Q$ relieves some of the representational issues that the director has, it is a more complicated object with its distinct order parameter space and topological classes. In the vicinity of a disclination core, nematic configurations become biaxial in three dimensions, and $\Q$ does not go to zero at the core \cite{schopohl87,schimming20b}. Results involving $\Q$ are presented below, but restricted to paths in real space in which the order parameter remains uniaxial. This is the case for distances away from the defect core larger than the coherence length of $\Q$.

Recent work introduced a tensor density $\mathbf{D}$, function of either the director $\n$ or the tensor order parameter $\mathbf{Q}$, which is nonzero near disclination cores. The tensor $\mathbf{D}$ was used not only to locate disclination lines \cite{schimming22}, but since it is related to the Jacobian of the transformation between real space and order parameter space, it led to a kinematic law relating the velocity of a disclination and the time derivative of the order parameter \cite{schimming23}. This law is independent of the dynamical model considered for $\Q$, and hence includes models of both passive and active nematics, as well as coupling to hydrodynamic flows. However, the precise connection between $\mathbf{D}$ and an invariant measure of disclination charge was not provided. We present an exact construction below.


Additionally, we explore cases in which $\mathbf{D}$ is nonzero in configurations that do not contain disclinations. For example, double-splay and double-twist nematic distortions generally give nonzero $\mathbf{D}$. We also extend our analysis to Skyrmions and point defects, both of which have had a long history of research and are of current interest for applications involving the design of unique meta-materials, colloidal assembly, and electroosmotic control of biomaterials \cite{stark01,lazo14,ackerman14,li17,peng18,duzgun21,duzgun22}. We show that $\mathbf{D}$ may be used to compute their topological charge, and to identify them, even as the structure of $\mathbf{D}$ differs substantially depending on the topological object under consideration.

\section{An invariant measure of disclination charge \label{sec:DiscCharge}}

We introduce a mathematical construction that generalizes Eqs. \eqref{eqn:PhiCharge}--\eqref{eqn:Qcharge} to three dimensions. The topological charge of a disclination line is always $+1/2$. We thus seek to construct a path integral in order parameter space that equals either $+1/2$ or zero, modulo $2\pi$, for any corresponding closed circuit in real space. We start with the quantity $\n\times d\n$ which gives the rotation of the director about the axis projected onto it \cite{efrati14}. Hence, the corresponding element of length of a curve in order parameter space is $ds^{2} = |\n \times d \n|^{2}$. For any locally defined unit vector $\XI$, the quantity $\XI \cdot \left(\n\times d\n\right)$ is the \textit{signed} projected element of length along the direction perpendicular to $\XI$ on the hemisphere. Given an arbitrary path on the unit hemisphere, parameterized by the arc length $s$, we first introduce a reference point on the path defined as $\n(0) = \n^*$. Then at each point $s$ on the path one defines,
\begin{equation} \label{eqn:XiDef}
    \XI(s) = \frac{\n(s) \times \T_{GC}(s)}{|\n(s) \times \T_{GC}(s)|} .
\end{equation}
Here $\T_{GC}(s)$ is the tangent vector to the great circle defined by the current point $\n(s)$ and the fixed point $\n^*$. Projecting this vector into $\n \times d\n$ gives
\begin{equation} \label{eqn:SigDef}
    \XI(s) \cdot \left(\n(s) \times d \n(s)\right) = \frac{\T_{GC}(s) \cdot d\n(s)}{|\n(s) \times \T_{GC}(s)|} = d\varsigma(s)
\end{equation}
where we have introduced the notation $\varsigma(s)$ to indicate the projected arc length along each great circle.

In order to construct the required succession of great circles on the unit sphere, we define the unit vector
\begin{equation}
    \mathbf{\hat{V}}(s) = a(s) \n^* + b(s)\n(s)
\end{equation}
where $a(s)$ and $b(s)$ are determined by requiring that $\mathbf{\hat{V}}(s)\cdot \n^* = 0$. Given, $\n^*$ and $\n(s)$, this can always be achieved via the Gram-Schmidt procedure. Then the curve
\begin{equation}
    \mathbf{\hat{W}}(t,s) = \cos t \n^* + \sin t \mathbf{\hat{V}}(s)
    \label{eq:Wdef}
\end{equation}
parameterizes the great circle on the unit sphere passing through both $\n^*$ and $\n(s)$ for any $s$. In particular, we have that $\mathbf{\hat{W}}(t^*(s),s) = \n(s)$ for $t^*(s) = -\arctan\left[1/a(s)\right]$. With this parameterization of the great circle, Eq. \eqref{eqn:SigDef} may be written as
\begin{equation}
    d\varsigma(s) = \left.\frac{d\mathbf{\hat{W}}/dt \cdot d\n/ds}{|\n \times d\mathbf{\hat{W}}/dt|}\right|_{t = t^*(s)} ds.
\end{equation}
Explicitly substituting the definition \eqref{eq:Wdef}, and integrating over a closed circuit, one has 
\begin{equation} \label{eqn:IntOfSigma}
    \oint_C d\varsigma(s) = \oint_C \frac{\hat{\mathbf{n}}(0) \cdot d \hat{\mathbf{n}}/d s}{|\hat{\mathbf{n}}(0) \times \hat{\mathbf{n}}(s)|} d s = \oint_C \frac{\n^* \cdot d\n}{|\n^* \times \n|}. 
\end{equation}
This is our central result. Note that 
\begin{equation}
    \nabla \times \frac{\n^*}{|\n^* \times \n|} = 0,
    \label{eq:curl}
\end{equation}
a result that can be proved by direct substitution in spherical coordinates. Therefore $d\varsigma$ is an exact differential on the hemisphere.

If the curve $C$ in Eq. \eqref{eqn:IntOfSigma} does not cross the equator of the unit sphere, the integral is zero because of Eq. (\ref{eq:curl}) and the fact that the curve starts and ends at the same point. On the other hand, if the curve does cross the equator then, since the integral is independent of the path, it will be equal to $\pi$ as this is the result for the shortest path joining the two points. In general, $\pi$ must be added to the contour integral for each time the equator is crossed, because, this is the length of the arc of a great circle connecting the two identified points on the equator. A subtle point here is that if the equator is passed an even number of times the measured configuration in three dimensions is topologically equivalent to a configuration with no disclinations. This is not represented by our measure since we are representing the ground state manifold ($\mathbb{RP}^2$) with vectors, and so we must impose that the resulting calculation is valid modulo $2 \pi$
\begin{equation} \label{eqn:3DDiscCharge}
    \oint_C d\varsigma(s) \in \left\{0,\frac{1}{2}\right\}\,\, \text{modulo}\,\, 2\pi.
\end{equation}

\begin{figure*}
    \centering
    \includegraphics[width = \textwidth]{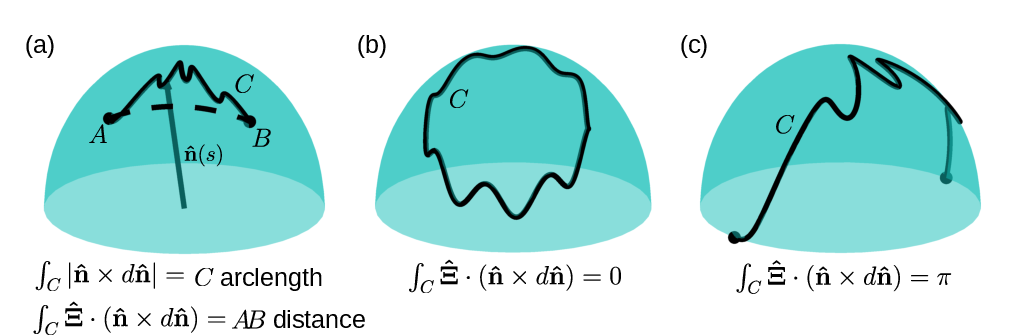}
    \caption{Illustration of curves in nematic order parameter space and corresponding path integrals. (a) Curve $C$ connecting two points $A$ and $B$. $\int_C |\n\times d\n|$ gives the arclength of contour $C$ while $\int_C \XI \cdot (\n \times d \n)$, with $\XI$ given by Eq. \eqref{eqn:XiDef}, gives the geodesic distance between $A$ and $B$. The dashed line shows the segment of great circle defined by $A$ and $B$, the length of which gives the shortest distance between the two. (b) A closed curve $C$ which does not pass through the equator. In this case $\int_C \XI \cdot (\n \times d\n) = 0$. (c) A closed curve $C$ which passes through the equator. In this case $\int_C \XI \cdot (\n \times d\n) = \pi$.}
    \label{fig:OPCurves}
\end{figure*}

The construction is graphically illustrated in Fig. \ref{fig:OPCurves} including the cases $\oint_C d\varsigma(s)$ equal zero or $\pi$ depending on whether the curve does or does not not pass through the equator. Note that the curves in both Figs. \ref{fig:OPCurves}b and \ref{fig:OPCurves}c are considered closed curves in the nematic order parameter space.

We may summarize our result for the charge of a disclination in the following way in terms of the measuring circuit in real space,
\begin{align} \label{eqn:3DQCharge}
    m &= \oint_C \hat{\Xi}_{\gamma} \varepsilon_{\gamma \mu \nu} \hat{n}_{\mu} \partial_k \hat{n}_{\nu} \, d \ell_{k} \,\, \text{modulo}\, \, 2 \pi \\
    &= \frac{1}{S_N^2} \oint_{C^*} \hat{\Xi}_{\gamma} \varepsilon_{\gamma \mu \nu} Q_{\mu \alpha} \partial_k Q_{\nu \alpha}\, d \ell_{k} \,\, \text{modulo} \, \, 2 \pi \nonumber
\end{align}
where we have used the identity $\n \times \nabla \n = S_N^2 \Q \times \nabla \Q$ for a uniaxial nematic with constant $S = S_N$. These results are independent of the circuit chosen.

It is interesting to note that as a measuring curve $C$ encircling a disclination line is taken to be smaller and smaller, the resulting curve on the ground state manifold itself approaches a great circle. In this limit, the vector $\XI$ becomes identical to the rotation vector $\OMEGA$, which describes the plane in which the director lies close to the disclination line in its normal plane.

\subsection{The disclination density tensor \bf{D}}

Equation \eqref{eqn:3DQCharge} may be used to locate disclination lines in a configuration. In practice, however, it is complicated to compute $\XI$ for various circuits $C$. Additionally, many curves must be constructed to completely locate the line. Therefore, just as is done in two dimensions, we construct a density by applying Stokes' theorem to Eq. \eqref{eqn:3DQCharge}. The curl of the integrand of Eq. \eqref{eqn:3DQCharge} is given by
\begin{widetext}
\begin{align}
    \varepsilon_{i k \ell} \partial_k \left(\hat{\Xi}_{\gamma} \varepsilon_{\gamma \mu \nu} \hat{n}_{\mu} \partial_{\ell} \hat{n}_{\nu}\right) &= \varepsilon_{i k \ell}\varepsilon_{\gamma \mu \nu} \left[\partial_k \hat{\Xi}_{\gamma} \hat{n}_{\mu} \partial_{\ell} \hat{n}_{\nu} + \hat{\Xi}_{\gamma} \partial_k \hat{n}_{\mu} \partial_{\ell} \hat{n}_{\nu} + \hat{\Xi}_{\gamma} \hat{n}_{\mu} \partial_k \partial_{\ell} \hat{n}_{\nu}\right] \label{eqn:nCurl} \\
    \varepsilon_{i k \ell} \partial_k \left(\hat{\Xi}_{\gamma} \varepsilon_{\gamma \mu \nu} Q_{\mu \alpha} \partial_{\ell} Q_{\nu \alpha}\right) &= \varepsilon_{i k \ell} \varepsilon_{\gamma \mu \nu} \left[\partial_k \hat{\Xi}_{\gamma} Q_{\mu \alpha} \partial_{\ell} Q_{\nu \alpha} + \hat{\Xi}_{\gamma} \partial_k Q_{\mu \alpha} \partial_{\ell} Q_{\nu \alpha} + \hat{\Xi}_{\gamma} Q_{\mu \alpha} \partial_k \partial_{\ell} Q_{\nu \alpha}\right] \label{eqn:QCurl}
\end{align}
\end{widetext}
where the derivative of $\XI$ may be computed by extending its definition to a family of curves that cover the surface of integration. In general there are three terms that must be integrated over the Stokes surface when computing the charge. If we consider the integrand near a disclination line, however, the expression simplifies significantly.

Near a disclination line, $\XI \to \OMEGA$ and an oriented orthonormal triad can be introduced $\{ \n_{0}, \n_{1}, \OMEGA \} $ so that the director is given by
\begin{equation} \label{eqn:nDisc}
\n = \n_{0} \cos \frac{\varphi}{2} + \n_{1} \sin \frac{\varphi}{2},
\end{equation}
where $\varphi$ is the azimuthal angle on the plane normal to the disclination tangent vector $\T$. Direct substitution of Eq. \eqref{eqn:nDisc} into Eq. \eqref{eqn:nCurl} gives $\varepsilon_{\gamma \mu \nu}\hat{n}_{\mu}\partial_{\ell}\hat{n}_{\nu} = (1/2)\hat{\Omega}_{\gamma} \partial_{\ell} \varphi$ and so the first term on the left hand side of Eq. \eqref{eqn:nCurl} goes to zero since $\hat{\Omega}_{\gamma}\partial_k \hat{\Omega}_{\gamma} = 0$, while the second term also goes to zero since $\varepsilon_{ik\ell}\varepsilon_{\gamma \mu \nu} \partial_k \hat{n}_{\mu} \partial_{\ell} \hat{n}_{\nu} = 0$. This leaves the third term as the only nonzero term for disclinations which gives a delta function due to the singular nature of the director at the core of the disclination. We note that there are nematic configurations in which the term $\varepsilon_{ik\ell}\varepsilon_{\gamma \mu \nu} \partial_k \hat{n}_{\mu} \partial_{\ell} \hat{n}_{\nu}$ is nonzero. We explore a few of these in the next section.

On the other hand, if the $\Q$ tensor representation is used and Eq. \eqref{eqn:QCurl} applies, a linear core approximation may be used for $\Q$ near the core \cite{long21,schimming22}:
\begin{multline} \label{eqn:LinearCore}
    \Q \approx S_N \left[ \frac{1}{6} \mathbf{I} - \frac{1}{2}\OMEGA \otimes \OMEGA  + \frac{\bm{\hat{\nu}}_0 \cdot \mathbf{r}}{2a}\left(\n_0 \otimes \n_0 - \n_1 \otimes \n_1\right) \right. \\ \left. + \frac{\bm{\hat{\nu}}_1 \cdot \mathbf{r}}{2 a} \left(\n_0 \otimes \n_1 + \n_1 \otimes \n_0\right)\right]
\end{multline}
where $\{\bm{\hat{\nu}}_0,\bm{\hat{\nu}}_1,\T\}$ are an orthonormal triad describing the orientation of the disclination line, and $a$ is the radius of the disclination core. Substituting Eq. \eqref{eqn:LinearCore} into Eq. \eqref{eqn:QCurl} gives $\varepsilon_{\gamma \mu \nu}Q_{\mu \alpha}\partial_{\ell}Q_{\nu \alpha} = \hat{\Omega}_{\gamma} A_{\ell}$ where 
$$
\mathbf{A} = \frac{\bm{\hat{\nu}}_0 \cdot \mathbf{r}}{2 a^2} \bm{\hat{\nu}}_1 - \frac{\bm{\hat{\nu}}_1 \cdot\mathbf{r}}{2 a^2} \bm{\hat{\nu}}_0. 
$$
Thus, the first term on the left hand side of Eq. \eqref{eqn:QCurl} is zero when $\XI \to \OMEGA$ for the same reason as above. Further, the third term in Eq. \eqref{eqn:QCurl} is always zero, regardless of the nematic configuration, since $\Q$ is a nonsingular quantity. This leaves the second term on the left hand side of Eq. \eqref{eqn:QCurl} as the only nonzero term.

The simplification of Eqs. \eqref{eqn:nCurl} and \eqref{eqn:QCurl} near a disclination core lead to the following definitions:
\begin{eqnarray} 
    \XI \cdot \mathbf{D}^{(\n)} & := & \hat{\Xi}_{\gamma} \varepsilon_{ik\ell} \partial_k \left(\varepsilon_{\gamma \mu \nu} \hat{n}_{\mu}\partial_{\ell} \hat{n}_{\nu}\right) 
\label{eqn:Dndef}
    \\
    \XI \cdot \mathbf{D}^{(\Q)} & := & \hat{\Xi}_{\gamma} \varepsilon_{\gamma \mu \nu} \varepsilon_{i k \ell} \partial_k Q_{\mu \alpha} \partial_{\ell} Q_{\nu \alpha}.
    \label{eqn:DQdef}
\end{eqnarray}
The tensor field $\mathbf{D}$, written in terms of either $\n$ or $\Q$, thus contains spatial information about disclinations. Its definition here coincides with Eq. (9) of Ref. \cite{schimming22}, in which the properties of $\mathbf{D}$ were explored for various disclination configurations. We note that the definition of $\mathbf{D}^{(\n)}$ is written to include both the second and third terms on the left hand side of Eq. \eqref{eqn:nCurl} so that $\mathbf{D}^{(\n)}$ may be used for both Skyrmions and point defects, as explored in the next section.

It is important to point out that the first term on the left hand side of Eqs. \eqref{eqn:nCurl} and \eqref{eqn:QCurl}, while zero at disclination cores, is not zero in general. As a consequence, a nonzero density $\mathbf{D}$ could point to a spurious topological singularity where there is none. Away from defects, where it is possible $\XI \neq \OMEGA$, this term may be nonzero. This is the case, for example, in double-splay or double-twist configurations \cite{selinger19,long21b}. Since a double-splay or double-twist configuration is not a disclination, the integral in Eq. \eqref{eqn:3DQCharge} must give zero, and so this term must also integrate to zero. That $\mathbf{D}$ itself is nonzero for these special configurations (and perhaps others) is an interesting result, and more work is needed to fully understand it. It is likely due to the fact that $\mathbf{D}$ is related to the Jacobian of the transformation between areas in configuration space and areas in subspaces of order parameter space. Real space patches of double-splay and double-twist configurations, for example, can be mapped to patches on the unit sphere. In the following section, we will explore some of the properties of $\mathbf{D}$ for nematic configurations containing double-splay or double-twist.

For the specific case of a line disclination, and by using either Eq. \eqref{eqn:nDisc} for $\n$ or Eq. \eqref{eqn:LinearCore} for $\Q$, it is possible to find a physically appealing decomposition of the $\mathbf{D}$ tensor. We find
\begin{equation} \label{eqn:DDecomp}
    \mathbf{D} = |\mathbf{D}|\left(\OMEGA \otimes \T\right)
\end{equation}
where $|\mathbf{D}^{(\n)}(\mathbf{r})| \propto \delta(\mathbf{r} - \mathbf{R})$ with $\mathbf{R}$ the location of the disclination core and $|\mathbf{D}^{(\Q)}(\mathbf{r})|$ is a diffuse scalar field with maximum at the disclination core \cite{schimming22}. Hence, the tensor $\mathbf{D}$ yields both the location and the geometric character of the disclination line from either $\n$ or the tensor $\mathbf{Q}$. This result is useful for both experiments and numerical computation to conveniently locate a disclination line and to determine its local geometric character.

\section{Properties of $\mathbf{D}$ for configurations not involving disclinations}

\subsection{Double-splay and double-twist}

As noted in the previous section, the tensor $\mathbf{D}$ may be nonzero in specific configurations that do not contain a topological defect, for example, in configurations with double-splay or double-twist distortion. We can explicitly calculate $\mathbf{D}$ in this case by considering $\n = \cos k \rho \mathbf{\hat{z}} + \sin k \rho \bm{\hat{\rho}}$ in cylindrical coordinates. This is an ideal ``double-splay'' configuration where the director is splayed in both directions, and $k$ characterizes the inverse length scale of the distortion. A double-twist configuration can be obtained by replacing $\bm{\hat{\rho}} \to \bm{\hat{\phi}}$ in the equation for $\n$. The tensor $\mathbf{D}$ at $\rho = 0$ is,
\begin{equation}
    \mathbf{D} = 2 k^2 \left(\mathbf{\hat{z}} \otimes \mathbf{\hat{z}}\right).
\end{equation}
We note that the result is the same for a double twist configuration. Thus, for these configurations $\mathbf{D} \neq 0$. 

On the other hand, computing the charge defined in Eq. \eqref{eqn:3DQCharge} by integrating along a curve around the double splay configuration will yield zero since the curve does not encircle a disclination. We may show this explicitly by computing the loop integral for a circle of radius $\rho$ with $k = \pi$. Using $\n^* = \cos \pi \rho \mathbf{\hat{z}} + \sin \pi \rho \mathbf{\hat{x}}$ the charge integral is
\begin{multline}
    -\tan \pi \rho \oint \frac{\sin \phi \, d\phi}{\sqrt{2 - 2\cos\phi + \sin^2\phi \tan^2\pi \rho}} = \\\begin{cases}
    0 & \rho \neq \frac{1}{2} \\
    -2 \pi & \rho = \frac{1}{2}
    \end{cases}
\end{multline}
which is always zero modulo $2 \pi$ as expected. The charge defined in Eq. \eqref{eqn:3DQCharge} along \textit{any} curve $C$ is zero in this configuration. This is because the corresponding curve in order parameter space will either not pass through the equator, or will do so an even number of times giving zero modulo two. As is well known, it is possible to continuously remove the distortion of this configuration to yield an undefected configuration. 

Despite the existence of regular configurations with nonzero $\mathbf{D}$, the tensor as defined is still useful to locate true singularities in both experiments and numerical computations since local relaxation can quickly remove large but regular distortions in systems in which the free energy penalizes them. Nevertheless, there are several liquid crystal systems which do support double-splay and double-twist distortions energetically. These are primarily cholesterics, in which the nematogens break chiral symmetry and in turn support spontaneous twist deformations \cite{deGennes75}. These systems have been shown to exhibit spontaneous double-twist regions separated by disclinations, known as \lq\lq blue phases'' \cite{meiboom82,wright89}. Additionally, and more recently, lyotropic chromonic liquid crystals have been shown to exhibit spontaneous double-twist configurations in confinement \cite{tortora11,davidson15,selinger22}. Further, topological defects other than disclinations may exhibit regions of double-splay and double-twist. For example, in liquid crystals that have a strong response to external fields, topological defects known as \lq\lq Skyrmions'' may form when an external field is introduced \cite{ackerman14,long21b,duzgun22}. These defects are not disclinations, and instead share properties with Skyrmions in magnetic systems \cite{fert17} and contain double-splay or double-twist configurations. We explicitly explore the properties of $\mathbf{D}$ for Skyrmions as well as point defects in the next sections. 

In systems in which elasticity or confinement promote double-splay or double-twist configurations, one would need to consider the director $\n$ or the tensor order parameter $\Q$ in addition to $\mathbf{D}$ to fully characterize regions where $\mathbf{D}$ is nonzero. This is not a problem, however, since $\mathbf{D}$ is computed from $\n$ or $\Q$ in the first place. Alternatively, formally, one may use the contour integral methodology as laid out above to unambigously identify the existence of a disclination.

\subsection{Skyrmions}

\begin{figure*}
    \centering
    \includegraphics[width = \textwidth]{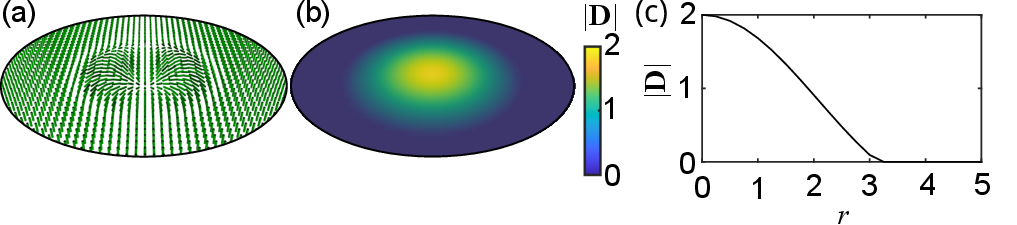}
    \caption{(a) Director field associated with a circular, splay-stabilized Skyrmion of radius $\pi$. (b) Spatial profile of $|\mathbf{D}|$ for the Skyrmion. (c) $|\mathbf{D}(\rho)|$ given by Eq. \eqref{eqn:DSkyrmion}.}
    \label{fig:Skyrmion}
\end{figure*}

Skyrmions are soliton-like topological defects that may occur in cholesterics or in liquid crystals with a strong response to external fields \cite{ackerman14,long21b,duzgun22}. Unlike disclinations, Skyrmions do not produce long ranged distortions of the nematic, and, instead, the nematic distortion is limited to a finite area, as shown in Fig. \ref{fig:Skyrmion}(a). Further, the Skyrmion nematic texture is not associated with a singularity in $\n$ and, thus, the tensor order parameter $\Q$ is not necessary to study the detailed structure of the defect. Despite these differences, the topological charge for a Skyrmion may still be defined via an integral,
\begin{equation} \label{eqn:nSkyrmion}
    m = \frac{1}{8 \pi} \int_{\Gamma} \hat{n}_{\gamma} \left(\varepsilon_{\gamma \mu \nu} \varepsilon_{i k \ell} \partial_k \hat{n}_{\mu} \partial_{\ell} \hat{n}_{\nu}\right) \, d\Sigma_{i}.
\end{equation}
Importantly, the surface $\Gamma$ that is integrated over is not closed and must ``cover'' the Skyrmion, otherwise a partial charge will be measured. We note that the quantity in parentheses is precisely the tensor $\DN$, defined in Eq. \eqref{eqn:Dndef}, such that there is no singularity in the director field. We may then rewrite Eq. \eqref{eqn:nSkyrmion} as 
\begin{equation} \label{eqn:nDCharge}
    m = \frac{1}{8 \pi} \int_{\Gamma} \hat{n}_{\gamma} D_{\gamma i}^{(\n)}\,d\Sigma_{i}.
\end{equation}

Similar to the case of disclinations, if one takes $\mathbf{Q} = S_N \left( \mathbf{\hat{n}} \otimes \mathbf{\hat{n}} - (1/3) \mathbf{I}\right)$ such that $S_N$ is constant, one can write
\begin{equation} \label{eqn:DCharge}
    m = \frac{1}{8 \pi S_N^2} \int_{\Gamma} \hat{n}_{\gamma} D_{\gamma i}^{(\Q)} \, d\Sigma_{i}.
\end{equation}
Since there is no singularity, $S$ should, in principle, be constant everywhere in the texture, so there is no difference in using the director representation versus the $\Q$ tensor representation. The integrands of Eqs. \eqref{eqn:nDCharge} and \eqref{eqn:DCharge} are similar to the quantity used to identify disclinations, Eqs. \eqref{eqn:Dndef} and \eqref{eqn:DQdef}, except that $\n$ is projected into $\mathbf{D}$ instead of $\XI$. This indicates that $\mathbf{D}$ may also be used to identify Skyrmions. 

As discussed above, for Skyrmions, $\DN$ will be nonzero due to the double-splay or double-twist configurations characteristic of these textures \cite{long21b}; however, the decomposition \eqref{eqn:DDecomp} will not hold for this type of distortion. We may calculate $\DN$ for an idealized, circular, splay-stabilized Skyrmion in which the director is given by $\n = \cos k \rho \mathbf{\hat{z}} + \sin k \rho \bm{\hat{\rho}}$ ($0 \leq \rho \leq  \pi / k$). We find
\begin{multline} \label{eqn:DSkyrmion}
    \DN = \frac{2 k \sin^2 k \rho}{\rho} \left(\bm{\hat{\rho}} \otimes \mathbf{\hat{z}}\right) + \frac{2 k \cos k\rho \sin k \rho}{\rho}\left( \mathbf{\hat{z}}\otimes \mathbf{\hat{z}}\right)
    \\ = \frac{2 k \sin k \rho}{\rho}\left(\n \otimes \mathbf{\hat{z}}\right)
\end{multline}
for $0 \leq \rho \leq \pi/k$ and $\DN = 0$ for $\rho > \pi /k$. In this case, $\DN$ has a diffuse, nonzero magnitude regardless of the representation of nematic. $|\mathbf{D}|$ is shown in Fig. \ref{fig:Skyrmion}(b,c) for an ideal, splay-stabilized Skyrmion. Further, while the decomposition \eqref{eqn:DDecomp} does not hold, $\DN \propto \left(\n \otimes \mathbf{\hat{N}}\right)$ where $\mathbf{\hat{N}}$ is the normal vector to the plane of the Skyrmion distortion.

\subsection{Point defects}

In three dimensional nematics, point defects are also topologically allowed. These objects manifest as point singularities in the director field, and, like disclinations, feature long ranged distortions of the nematic. The charge of a point defect may be measured by an integral similar to that of Skyrmions:
\begin{equation} \label{eqn:PointCharge}
    m = \frac{1}{8\pi}\oint_{\partial \Omega}\hat{n}_{\gamma}\left(\varepsilon_{\gamma \mu \nu} \varepsilon_{i k \ell} \partial_k\hat{n}_{\mu}\partial_{\ell}\hat{n}_\nu\right) \, d\Sigma_i
\end{equation}
where the surface of integration $\partial \Omega$ is a closed surface, and hence a boundary of a volume $\Omega$. As with Skyrmions, we may write this charge in terms of the tensor $\mathbf{D}$,
\begin{align}
    m &= \frac{1}{8\pi}\oint_{\partial \Omega} \hat{n}_{\gamma}D_{\gamma i}^{(\n)}\, d\Sigma_i \label{eqn:nPointCharge} \\
    m &= \frac{1}{8 \pi S_N^2}\oint_{\partial \Omega^*} \hat{n}_{\gamma}D_{\gamma i}^{(\Q)}\,d\Sigma_i \label{eqn:QPointCharge}
\end{align}
where $\partial \Omega^*$ denotes a surface such that the scalar order parameter remains constant, $S = S_N$.

The properties of $\mathbf{D}^{(\n)}$ and $\mathbf{D}^{(\Q)}$ for point defects are different because of the director singularity. As an example, we compute $\mathbf{D}^{(\n)}$ explicitly for an ideal ``radial hedgehog'' point defect with $\n = \mathbf{\hat{r}}$:
\begin{equation} \label{eqn:DHedge}
    \mathbf{D}^{(\n)} = \frac{2}{r^2}\left(\mathbf{\hat{r}}\otimes \mathbf{\hat{r}}\right) = \frac{2}{r^2}\left(\n \otimes \mathbf{\hat{r}}\right)
\end{equation}
where the second equality is written as a conjecture for a more general decomposition of $\mathbf{D}^{(\n)}$ for a point defect. Equation \eqref{eqn:DHedge} shows that the tensor $\mathbf{D}^{(\n)}$ diverges at $r = 0$ where the director singularity occurs. More interestingly, while $|\mathbf{D}^{(\n)}|$ decays rapidly, it does not go to zero at some finite distance. This behavior is strikingly different than in the case of a disclination, in which $|\mathbf{D}^{(\n)}| \propto \delta(\mathbf{r})$. 

The decomposition of $\mathbf{D}^{(\n)}$ for disclinations into tangent and rotation vectors does not hold for point defects, as can be seen by comparing Eqs. \eqref{eqn:DDecomp} and \eqref{eqn:DHedge}. We further probe the suggested decomposition in Eq. \eqref{eqn:DHedge} by computing $\mathbf{D}^{(\n)}$ for a negative hedgehog point defect in which $\n = - \cos2\theta \mathbf{\hat{r}} + \sin2\theta \bm{\hat{\theta}}$. In this case
\begin{equation}
    \mathbf{D}^{(\n)} = \frac{2}{r^2}\left[\cos2\theta \left(\mathbf{\hat{r}}\otimes\mathbf{\hat{r}}\right) - \sin2\theta \left(\bm{\hat{\theta}}\otimes\mathbf{\hat{r}}\right)\right] = -\frac{2}{r^2}\left(\n \otimes \mathbf{\hat{r}}\right)
\end{equation}
hence the charge of the defect is reflected in the sign of $\mathbf{D}^{(\n)}$, and the conjectured decomposition holds. We note that we have used here the convention that a positive hedgehog is one such that $\n = +\mathbf{\hat{r}}$ pointing outward. However, the sign of $\mathbf{D}^{(\n)}$ changes if $\n \to -\n$ and so too will the measured charge as reflected by Eq. \eqref{eqn:PointCharge}. This ``local-to-global'' problem is discussed in more detail in Ref. \cite{alexander12}.

\begin{figure*}
    \centering
    \includegraphics[width = \textwidth]{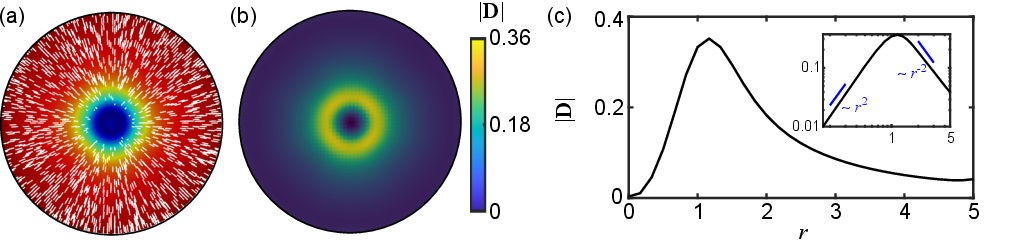}
    \caption{(a) Nematic profile of a cut through a radial hedgehog point defect, numerically computed using a $\Q$-tensor representation. The color indicates the scalar order parameter, $S$, with $S\to0$ at the defect core. The white lines show the local director $\n$. (b) Numerically computed spatial profile of $|\mathbf{D}^{(\Q)}|$ for the hedgehog point defect. (c) Plot of $|\mathbf{D}^{(\Q)}(r)|$ extracted from (b). Inset: Same plot on logarithmic scaling axes.}
    \label{fig:Hedgehog}
\end{figure*}

If we instead compute $\mathbf{D}^{(\Q)}$ for a point defect, there is no longer a divergence at the defect core. This is because $\Q \to 0$ at the defect core to alleviate the diverging elastic energy, as shown in Fig. \ref{fig:Hedgehog}(a) in which we have numerically computed the structure of a radial hedgehog defect by minimizing a free energy in terms of $\Q$ \cite{schimming21}. However, asymptotically $\Q \sim r^2$ at the core of a point defect \cite{greco92,majumdar12}, and so $\mathbf{D}^{(\Q)} \to 0$. Thus, apparently, the topological information is lost at the core of the point defect. We have computed $\mathbf{D}^{(\Q)}$ from the hedgehog profile in Fig. \ref{fig:Hedgehog}(a) which is shown in Figs. \ref{fig:Hedgehog}(b,c). We find $|\mathbf{D}^{(\Q)}| \sim r^2$ close to the defect core, while $|\mathbf{D}^{(\Q)}| \sim 1/r^2$ far from the defect core where $S$ is constant, as shown in the inset of Fig. \ref{fig:Hedgehog}(c). This behavior is a major difference from the case of disclinations in which $\mathbf{D}^{(\Q)}$ is finite within the core region, and largest at the disclination core. This apparent loss of topological information may be explained by the fact that the full order parameter space for $\Q$ is biaxial, and does not support point defects \cite{re:pismen99}. Hence a point defect in the $\Q$-tensor representation may dissociate into a biaxial disclination loop. Outside the core, when $S = S_N$ is constant, we still have $|\mathbf{D}^{(\Q)}| \sim 1/r^2$ and so Eq. \eqref{eqn:QPointCharge} still holds since we only consider surfaces in which $S = S_N$ throughout the surface.

Finally, because the surface integrated over in Eq. \eqref{eqn:PointCharge} is closed, one may ask if a topological charge density akin to $\mathbf{D}$ for disclinations may be defined. Applying Gauss' law to Eqs. \eqref{eqn:nPointCharge} and \eqref{eqn:QPointCharge} yields
\begin{align}
    m &= \frac{1}{8\pi}\int_{\Omega} \partial_i\left(\hat{n}_{\gamma}D_{\gamma i}^{(\n)}\right) \, dV \\
    m &= \frac{1}{8\pi S_N^2}\int_{\Omega^*} \partial_i\left(\hat{n}_{\gamma}D_{\gamma i}^{(\Q)}\right) \, dV.
\end{align}
The integrands of the above equations act as effective point defect densities. For the positive and negative hedgehogs computed above, the density is $\nabla \cdot \left(\n \cdot \mathbf{D}^{(\n)}\right) \propto \delta(r)$ as expected. On the other hand, $\nabla \cdot \left(\n \cdot \mathbf{D}^{(\Q)}\right)$ is a diffuse scalar field which is still zero at the core of the point defect, but nonzero in the region surrounding the core in which $\Q$ is varying. 

\section{Conclusions}

We have introduced an exact expression defining the topological charge of line defects encircled by an arbitrary path in $\mathbb{RP}^2$. It yields $+1/2$ for configurations with net disclination charge and zero otherwise, as required by the topology of disclinations in three-dimensional nematics. This method may be used to unambiguously identify disclinations given a path in order parameter space.

The integral over paths can be conveniently transformed into surface integrals involving the tensor density $\mathbf{D}$. While $\mathbf{D}$ is a useful quantity to identify and locate disclination lines from both director and tensor order paramter field, it can also be nonzero in certain configurations which do not contain disclinations. For example, double-splay and double-twist nematic configurations generically yield a nonzero $\mathbf{D}$, which results in a nonzero $\mathbf{D}$ for other topologically protected objects such as Skyrmions and point defects. For these objects, we have shown that $\mathbf{D}$ has different properties than for the case of disclinations, but that it may still be used to identify and potentially characterize them.

Elucidating the consequences of topological constraints on the properties and evolution of nonequilibrium soft matter systems remains a very active area of research. Our results provide the tools to diagnose the existence of disclination lines in experimentally determined or computationally generated configurations of nematics, as well as predicting and tracking their motion. Further research is needed to confirm the decomposition of $\mathbf{D}$ proposed for both Skyrmions and point defects as our results apply to specific, idealized, configurations, but we have not proven them for general cases. Additionally, we have focused here on $\mathbf{D}$ as defined for nematic phases but it appears possible that a similar tensor may be defined for other systems such as smectics or magnetic materials. It is already known that a similar object exists for dislocations in solids \cite{kleman08,skogvoll22}. Finally, as demonstrated in Ref. \cite{schimming23}, there is a direct connection between $\mathbf{D}$ and disclination dynamics. It remains to be examined whether a similar connection could be made to understand the dynamics of Skyrmions and point defects.

\begin{acknowledgements}
We are indebted to Jonathan Selinger for clarifying the scope of the definition of the tensor $\mathbf{D}$. His comments led to the redefinition of its path integral around a disclination core from the earlier version given in \cite{schimming22}. This research has been supported by the National Science Foundation under Grant No. DMR-1838977, and by the Minnesota Supercomputing Institute. C.D.S. acknowledges support from the U.S. Department of Energy through the Los Alamos National Laboratory. Los Alamos National Laboratory is operated by Triad National Security, LLC, for the National Nuclear Security Administration of U.S. Department of Energy (Contract No. 89233218CNA000001).
\end{acknowledgements}

\bibliography{LC}

\end{document}